\documentclass[runningheads]{llncs}
\usepackage[T1]{fontenc}
\usepackage{graphicx,verbatim}

\usepackage[pagebackref,breaklinks,colorlinks]{hyperref}

\input{preamble}
\begin{document}
\title{Conditional diffusion models for guided anomaly detection in brain images using fluid-driven anomaly randomization}
%

\author{Ana Lawry Aguila\inst{1}
\and
Peirong Liu\inst{1}
\and
Oula Puonti\inst{2}
\and
Juan Eugenio Iglesias\inst{1}
}
\authorrunning{A. Lawry Aguila et al.}
\institute{Athinoula A. Martinos Center for Biomedical Imaging, Massachusetts General Hospital and Harvard Medical School, Boston, USA \and Danish Research Centre for Magnetic Resonance, Department of Radiology and Nuclear Medicine, Copenhagen University Hospital– Amager and Hvidovre, Copenhagen, Denmark\\
\email{acaguila@mgh.harvard.edu}}

\maketitle              
\begin{abstract}
%
Supervised machine learning has enabled accurate pathology detection in brain MRI, but requires training data from diseased subjects that may not be readily available in some scenarios, for example, in the case of rare diseases. Reconstruction-based unsupervised anomaly detection, in particular using diffusion models, has gained popularity in the medical field as it allows for training on healthy images alone, eliminating the need for large disease-specific cohorts. These methods assume that a model trained on normal data cannot accurately represent or reconstruct anomalies. However, this assumption often fails with models failing to reconstruct healthy tissue or accurately reconstruct abnormal regions i.e., failing to remove anomalies. In this work, we introduce a novel conditional diffusion model framework for anomaly detection and healthy image reconstruction in brain MRI. Our weakly supervised approach integrates synthetically generated pseudo-pathology images into the modeling process to better guide the reconstruction of healthy images. To generate these pseudo-pathologies, we apply fluid-driven anomaly randomization to augment real pathology segmentation maps from an auxiliary dataset, ensuring that the synthetic anomalies are both realistic and anatomically coherent. We evaluate our model's ability to detect pathology, using both synthetic anomaly datasets and real pathology from the ATLAS dataset. In our extensive experiments, our model: \textit{(i)}~consistently outperforms variational autoencoders, and conditional and unconditional latent diffusion; and \textit{(ii)}~surpasses on most datasets, the performance of supervised inpainting methods with access to paired diseased/healthy images.

\end{abstract}
\section{Introduction}

Data availability presents a significant challenge in the application of supervised machine learning for anomaly detection in brain MRI, particularly when labels at the voxel level are desirable -- since manual annotation of pathology is time-intensive and requires specialized expertise. As a result, large-scale datasets with high-quality pathology annotations are limited. Unsupervised anomaly detection has gained popularity in the medical field as it enables training solely on healthy images, removing the need for such large disease cohorts or assumptions about the form anomalies take~\cite{zhou2020,pinaya2021,Jiang2023,schlegl2017,schlegl2019,zimmerer2018,chen2018,chen2020}. Reconstruction-based methods, such as autoencoders~\cite{Gong2019,zong2018,Baur2020,Denouden2018,Kumar2021}, assume that models trained on healthy data poorly reconstruct out-of-distribution (OoD) regions, and will instead reconstruct a closely matched healthy counterpart. Anomaly maps can be created by comparing the generated pseudo-healthy with the original disease images. The pseudo-healthy images or subsequent anomaly maps can be used for downstream tasks such as anomaly segmentation~\cite{Baur2020} or image inpainting~\cite{Durrer2024b}. However, several studies have found that autoencoders are unable to detect abnormality in several OoD samples~\cite{Bercea2023,Zhou2023} and suffer from poor generative ability, resulting in blurred reconstructions and high reconstruction errors even in healthy regions~\cite{Bercea2023}. 
Denoising Diffusion Probabilistic Models (DDPMs)~\cite{Ho2020} have recently been applied to anomaly detection~\cite{Pinaya2022,Bercea2024c,Bercea2023b,Graham2023b,graham2023c}, demonstrating significant improvements over autoencoder-based methods. Through a process of gradually adding and subsequently removing noise from an image, diffusion models are able to capture complex data distributions. In unsupervised anomaly detection, they function similarly to other reconstruction-based methods, training on healthy images with the assumption that denoising a noised diseased image will replace anomalous regions with healthy tissue. However, particularly when large anomalous regions are present, it can be difficult to select a noise level that effectively inpaints anomalies without removing distinctive features of healthy tissue.

Weakly, or self-supervised learning using synthetic anomalies offer a promising alternative to unsupervised approaches to pathology detection and pseudo-healthy image generation. Incorporating both healthy and diseased images into the training process enables the model to learn a more diverse data distribution, potentially improving its ability to guide the image reconstruction process. Recently, several methods have been proposed for healthy image generation and anomaly detection in brain MRI using synthetic data~\cite{Liu2025,Baugh2024,laso2024quantifying,liu2024pepsi,Liu_2023_BrainID}. One representative method is the recently proposed UNA~\cite{Liu2025}, which addresses the scarcity of gold-standard pathology segmentations by using fluid-driven anomaly randomization to generate realistic brain pathology profiles on-the-fly, which are then used for reconstructing healthy brain images. However, UNA tends to “over-correct” its reconstructed healthy anatomy, particularly in cases where severe pathology in the input image significantly obscures the underlying anatomy. A recent work~\cite{Baugh2024} incorporates synthetic pathology into a 2D diffusion model. However, the absence of real, manually annotated pathology masks in their anomaly generation framework may limit the realism of the generated anomalies.

In this work, we present the first diffusion model-based method for anomaly detection in 3D brain MRI which incorporates realistic synthetic pathology during training. Our model improves the anomaly detection ability of DDPMs by incorporating pseudo-disease images, via an auxiliary encoder network. These images are generated for each healthy training counterpart using a fluid-driven anomaly randomization approach which augments existing gold-standard pathology segmentations to generate realistic synthetic pathology that guides the reconstruction process. We train our model in a weakly supervised fashion where we condition the information derived from pseudo-pathology images at each timestep of the reverse diffusion process. This training approach allows our model to correct synthetic anomalies during training whilst preserving regions of healthy tissue. Our method achieves state-of-the-art anomaly detection across synthetic datasets and real pathology from the ATLAS dataset, showcasing the efficacy of our model in detecting anomalies in 3D brain images. 



\section{Methods}

\subsection{Latent diffusion models}\label{Sec:Latent Diffusion Models}

In this work, we build upon the Latent Diffusion Model (LDM) \cite{Rombach2021} framework, which is trained in two stages. In the first stage, an encoder $E_{\sigma}(\cdot)$ maps an input image $\mathbf{x} \in \mathbb{R}^{H \times W \times D \times 1}$ to a latent representation $\mathbf{z}_0 \in \mathbb{R}^{h \times w \times d \times c}$, and a decoder $D_{\phi}(\cdot)$ reconstructs $\mathbf{x}$ from $\mathbf{z}_0$. In the second stage, a DDPM~\cite{Ho2020} is trained to learn the distribution of $\mathbf{z}_0$. The diffusion process has two components. Firstly, the forward noising process consists of a Markov chain which progressively adds Gaussian noise to $\mathbf{z}_0$ such that after $T$ noising steps $\mathbf{z}_T \sim \mathcal{N}(0, \mathbf{I})$. Each step is defined as $q(\mathbf{z}_t|\mathbf{z}_{t-1}) := \mathcal{N}(\mathbf{z}_t; \sqrt{\alpha_t} \mathbf{z}_{t-1}, (1-\alpha_t)\mathbf{I})$ where $\alpha_t$ controls the noise level at step $t$. Secondly, the denoising reverse processes, another Markov chain, learns to progressively denoise $\mathbf{z}_T$ where $p_\theta\left(\mathbf{z}_{t-1} \mid \mathbf{z}_t\right)=\mathcal{N}\left(\mu_\theta\left(\mathbf{z}_t\right), \Sigma_\theta\left(\mathbf{z}_t\right)\right)$
with the variational lower bound on the log-likelihood of our latent $\mathbf{z}_{0}$ reducing to:
\begin{equation}
\begin{aligned}
    \label{eq:lower-bound}
    \mathbb{E} \left[ \log p_{\theta}(\mathbf{z}_{0}) \right] & \geq 
    \log p\left(\mathbf{z}_0 | \mathbf{z}_1\right)  - \sum_t D_{KL}\left(q\left(\mathbf{z}_{t-1} | \mathbf{z}_t, \mathbf{z}_0\right) \| p_\theta\left(\mathbf{z}_{t-1} | \mathbf{z}_t\right)\right) \,.
\end{aligned}
\end{equation}
Following the parametrization from~\cite{Ho2020}, $\mu_\theta$ can be modeled using a denoising model $\epsilon_\theta$ which can be trained with the simple objective:
\begin{equation}
\label{eq:obj}
\mathcal{L}=\mathbb{E}_{\mathbf{z}_{0} \sim q(\mathbf{z}_{0}), \epsilon_{t} \sim \mathcal{N}(\mathbf{0}, \mathbf{I})}\left[\left\|\epsilon_t-\epsilon_\theta\left(\mathbf{z}_t, t\right)\right\|_2^2\right] \,.
\end{equation}
To train diffusion models with a learned reverse process covariance $\Sigma_\theta$ it is necessary to optimize the full loss in Equation \ref{eq:lower-bound}. $\Sigma_\theta$ is implemented using $\epsilon_\theta$. 

\subsection{Our proposed framework}\label{Sec: Our proposed framework}

Inspired by UNA~\cite{Liu2025}, we use fluid-driven anomaly randomization to generate pseudo-pathological images that guide our model training. This method frames anomaly pattern randomization as an advection-diffusion process governed by partial differential equations (PDEs), ensuring well-posed simulated anomalies through controllable velocity fields and boundary conditions.  

Let $P({\mathbf{x}},\, t)$ denote the pathology probability of ${\mathbf{x}}$ at time $t$ where $x \in \Omega$ and $\Omega \subset \mathbb{R}^3$. The following PDE describes the anomaly randomization process:
\begin{equation}
\frac{\partial P({\mathbf{x}},\, t)}{\partial t} = - \nabla \times \boldsymbol{\Psi}({\mathbf{x}})\cdot\nabla P({\mathbf{x}}, t) + \nabla \cdot \left({\Phi}^2({\mathbf{x}})\, \nabla P({\mathbf{x}}, t)\right) \,,
\end{equation}
\begin{equation}
\text{with constraints:} \quad P({\mathbf{x}}, \, 0) = P_0({\mathbf{x}}), \quad \frac{\partial P({\mathbf{x}}, \, t)}{\partial \mathbf{n}}\big|_{\partial \Omega} = 0, \quad t \leq T_{\text{max}},\,
\end{equation}where $t$ ($T_{\text{max}}$) refers to the (maximum) integration time for anomaly randomization. $\boldsymbol{\Psi}$ and $\Phi$ represent potentials for the velocity and diffusion fields which determine the advection and diffusion process of the initial anomaly $P_0({\mathbf{x}})$. These potentials ensure incompressible flow and non-negative diffusion. The second constraint on anomaly boundaries ($\frac{\partial P({\mathbf{x}}, \, t)}{\partial \mathbf{n}}\big|_{\partial \Omega}$) ensures that the randomization process of $P_0$ stays within the defined bounds, i.e. the brain. Following UNA~\cite{Liu2025}, the initial pathology probability $P_0$ is generated on-the-fly using realistic pathology segmentation maps. $P_0$ is then randomized with random potentials $\boldsymbol{\Psi}$ and $\Phi$ derived from Perlin noise. Anomalies are encoded into a healthy image $\mathbf{x}_h$:
\begin{equation}
    \mathbf{x}_p = \mathbf{x}_h + \Delta \mathbf{x}_h \cdot P_{T_{\text{max}}}({\mathbf{x}})\,,
\end{equation}
where $\Delta \mathbf{x}_h$ depends on the mean of white matter intensities ($\mu_{\text{w}}$) in $P_0$:
\begin{equation}
\Delta \mathbf{x}_h \sim 
\begin{cases} 
0 & \text{if } {\mathbf{x}} \notin \Omega_{P}\,, \\
\mathcal{N}(-\mu_{\text{w}} / 2 ,\, \mu_{\text{w}} / 2 ) & \text{if } {\mathbf{x}} \in \Omega_{P}\,.
\end{cases}    
\end{equation}
based on prior knowledge of white and gray matter intensities in T1-weighted MRI. To simulate extreme scenarios, we randomly flip the sign of $\Delta  \mathbf{x}_h$ by 20\%. 

\begin{figure*}[t]
    \centering
    \includegraphics[trim={0 0.2cm 0 0}, clip, width=\columnwidth]{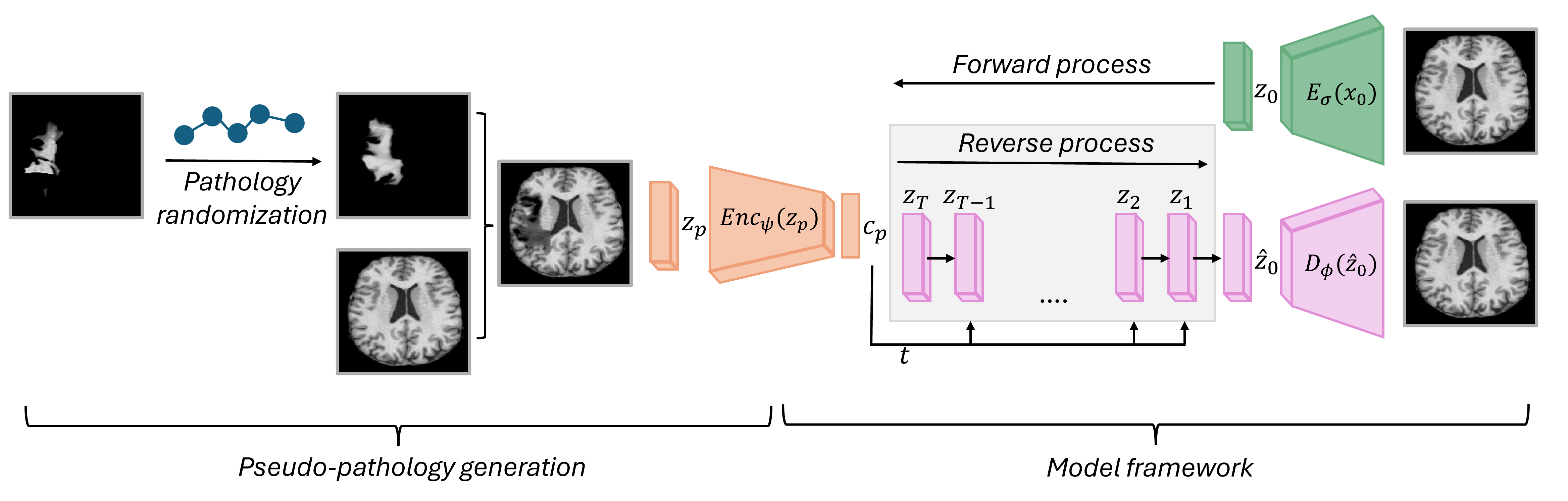}
    \caption{The left side of the figure depicts the pathology encoding process, where fluid-driven randomization generates a realistic synthetic pseudo-pathology image. The right side shows the conditional latent diffusion model, which is trained using both the original healthy image and the synthetic pseudo-pathology image.}
    \label{fig:CADD_framework}
\end{figure*}

To guide the healthy image reconstruction process, we incorporate the pseudo-pathology image counterparts via conditioning such that Equation \ref{eq:obj} becomes:
\begin{equation}
\label{eq:obj_c}
\mathcal{L}=\mathbb{E}_{\mathbf{z}_{0} \sim q(\mathbf{z}_{0}), \epsilon_{t} \sim \mathcal{N}(\mathbf{0}, \mathbf{I})}\left[\left\|\epsilon_t-\epsilon_\theta\left(\mathbf{z}_t,\mathbf{c}_p, t\right)\right\|_2^2\right]\,,
\end{equation}
where $\mathbf{c}_p$ encodes information from the pseudo-pathology image embedding, $\mathbf{z}_p$. To learn $\mathbf{c}_p$, we train an auxillary network, $\text{Enc}_{\varphi}(\cdot)$, to capture the most meaningful information from $\mathbf{z}_p$. We implement $\epsilon_\theta$ as a UNet~\cite{Ronneberger2015,Rombach2021,Pinaya2022,Graham2023b} and $\text{Enc}_{\varphi}(\cdot)$ has the same architecture as the encoder of the UNet~\cite{Preechakul2022}. Following~\cite{Preechakul2022}, we integrate timestep $t$ and pathology information $\mathbf{c}_p$ into our model using adaptive group normalization layers (AdaGN)~\cite{Dhariwal2021} which are applied throughout the UNet, ensuring strong conditioning of pathology information within the network.

For anomaly detection, we use the predicted pseudo-healthy reconstruction $\mathbf{x}_0^T=D_{\phi}(\mathbf{z}_0^T)$ and the original image $\mathbf{x}_0$ to generate anomaly maps, $m$. We allow a maximum 2 pixel-wise shift in each direction in the reconstructed image, to adjust for slight misalignment of the pseudo-healthy image, minimizing the absolute difference between the original and pseudo-healthy reconstruction. We weight the pixel-wise mean absolute difference by a measure of whole image similarity, min-max normalize to [0,1], and apply a median filter with kernel size 5 and brain mask eroding for 6 iterations, $cd$, such that~\cite{Bercea2024c}: 
\begin{equation}\label{eq:anomaly_map}
    m(\mathbf{x}_0, \mathbf{x}_0^T) = cd(\left|\mathbf{x}_0-\mathbf{x}_0^T\right| \cdot \text{LPIPS}_{\text{Alex}}\left(\mathbf{x}_0, \mathbf{x}_0^T\right)) \,,
\end{equation}
where $\text{LPIPS}_{\text{Alex}}$ is the AlexNet~\cite{Krizhevsky2012} learned perceptual image patch similarity.

\section{Experimental setup}
\noindent\textbf{Datasets.} To train our method, we use subjects from the following datasets; ADNI~\cite{Weiner2017} ($N_{\text{train}}$=270, $N_{\text{val}}$=15, $N_{\text{test}}$=31), HCP~\cite{Essen2012} ($N_{\text{train}}$=701, $N_{\text{val}}$=38, $N_{\text{test}}$=82), ADHD200~\cite{Brown2012} ($N_{\text{train}}$=706, $N_{\text{val}}$=32, $N_{\text{test}}$=82), AIBL~\cite{Fowler2021} ($N_{\text{train}}$=545, $N_{\text{val}}$=34, $N_{\text{test}}$=64), and OASIS3~\cite{LaMontagne2018} ($N_{\text{train}}$=1057, $N_{\text{val}}$=53, $N_{\text{test}}$=123). We use manual lesion segmentation maps from subjects from the ATLAS dataset~\cite{Liew2017} ($N_{\text{train}}$=590, $N_{\text{test}}$=56) to generate pseudo-pathology images. Healthy scans are augmented to encode synthetic anomalies where the synthetic abnormal profiles are
generated using fluid-driven anomaly randomization as described in Section \ref{Sec: Our proposed framework}, with initial profiles sampled from the lesion segmentation maps of ATLAS subjects. We encode anomalies into 80\% of training samples. 

For all datasets, we use T1-weighted MRI scans which have been skull stripped and bias field corrected with FreeSurfer~\cite{Fischl2012}, and min-max normalized to [0,1]. Affine registration was used to register images to the MNI 152 brain template using pre-computed (using EasyReg~\cite{Iglesias2023b,Hoffmann2022}) affine matrices applied during pre-processing. Images were cropped to 160 x 160 x 160.

\noindent\textbf{Implementation details.} As the first stage, we use an autoencoder with a KL-regularised latent space and perceptual and patch-based adversarial objectives~\cite{Rombach2021} which maps the brain image to a latent representation of size $3 \times 20 \times 20 \times 20$. We use the training parameters given by~\cite{pinaya2022b}. For the second stage, our model uses the conditional diffusion model backbone described in Section \ref{Sec: Our proposed framework}. We adopt the UNet architecture from~\cite{Graham2023b,Rombach2021}, which features three levels with channel configurations of (128, 256, 256). Each level incorporates a residual block, while an attention mechanism is applied at the deepest level. The condition vector, $\mathbf{c}_p = \text{Enc}_{\varphi}(\mathbf{z}_p)$, is a flattened $d$ dimensional vector where $d=1280$. Timesteps are sinusoidally embedded and processed through a two-layer MLP with Swish activation~\cite{ramachandran2017}. During training, we use $T = 1000$ and apply a linear noise schedule with $\beta_t$ ranging from 0.0015 to 0.0195. All models are trained using the Adam optimiser~\cite{kingma2017} with early stopping criteria on the validation loss and a learning rate of 0.0001. We use $T_\text{int}=250$ at inference time for anomaly detection, as done in previous works~\cite{Bercea2023b,Bercea2024c}. Random seeds are set at inference time for reproducibility. We use the anomaly randomization parameters given by~\cite{Liu2025} for generating pseudo-pathology. To ease the computational burden of our model, we pre-compute 8 synthetic pathology images for each sample of the training and validation cohorts and 2 images for the test cohorts.

\noindent\textbf{Comparison methods.} We compare our model with the following unsupervised generative-modelling-based anomaly detection approaches: a variational autoencoder (VAE)~\cite{Baur2020}, LDM~\cite{Rombach2021}, LDM ($T_{\text{avg}}$)~\cite{Graham2023b}. Where possible and available, we use the code from the original implementation. We use the same first-stage model for LDM baselines. We also compare with a conditional LDM (cLDM) where we concatenate $\mathbf{z}_p$ to $\mathbf{z}_t$ as additional channel inputs to $\epsilon_\theta$, in a similar approach to~\cite{Baugh2024}. Finally, we also compare our model with two fully supervised approaches; SynthSR~\cite{Iglesias2023}, a machine learning method for joint super-resolution, T1 contrast synthesis, and anomaly inpainting of brain MRI scans, and UNA~\cite{Liu2025}, a general-purpose model for diseased-to-healthy image generation. 
Note that as SynthSR reconstructs the skull, we use SynthSeg~\cite{Billot2023} to skull strip its output images.

\noindent\textbf{Comparison metrics.} 
From anomaly maps (Equation \ref{eq:anomaly_map}), we calculate the average pixel-wise area under the ROC curve ($AUC_{\text{pix}}$), pixel-level average precision ($AP_{\text{pix}}$), maximum Dice index across all thresholds per sample, and the false positive rate (FPR) based on Dice index threshold. We evaluate on both simulated pathology and real pathology from the ATLAS test set. In both scenarios, the model takes only pathology samples as input, using them for $\mathbf{z}_0^T$ and $\mathbf{z}_p$.

\begin{figure*}[th]
    \centering
    \begin{subfigure}{0.24\linewidth} 
        \centering
        \includegraphics[width=0.5\linewidth, trim=5 5 480 0, clip]{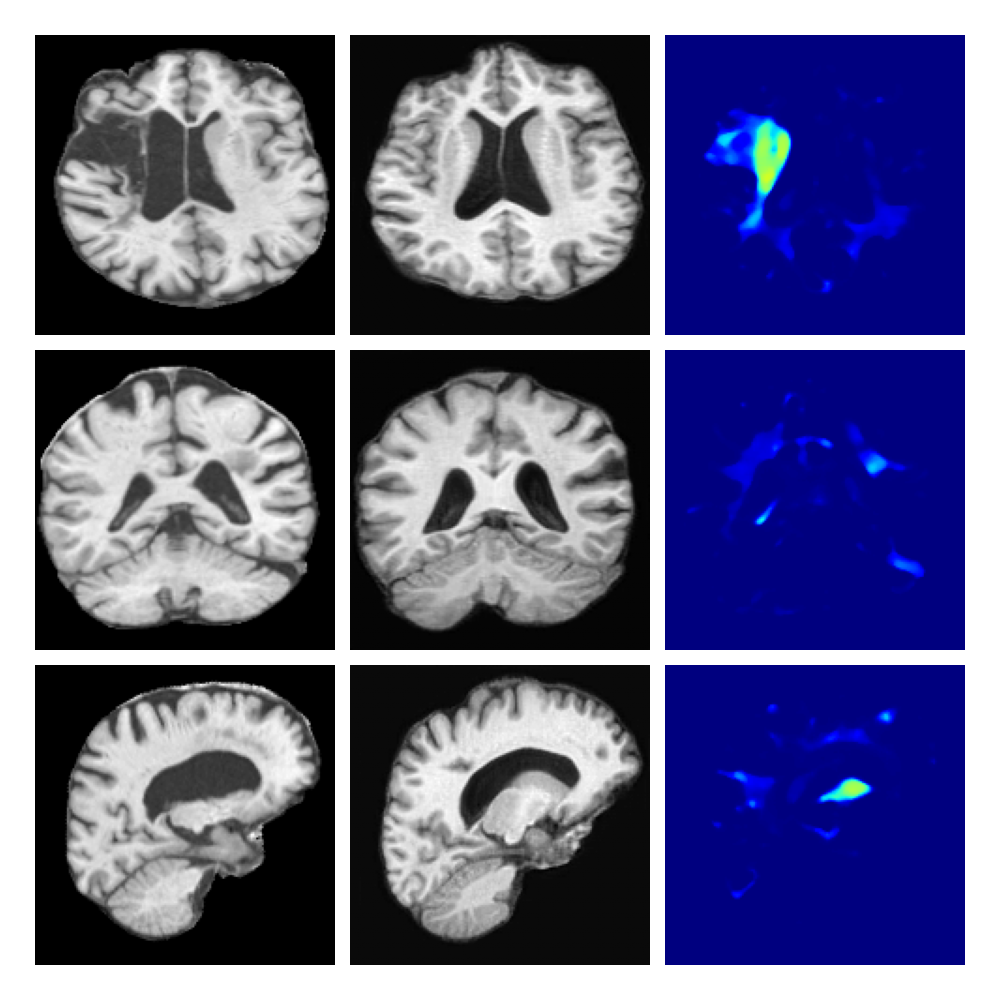}
        \caption{Original}
    \end{subfigure}
    \begin{subfigure}{0.24\linewidth} 
        \centering
        \includegraphics[width=\linewidth, trim=245 5 5 0, clip]{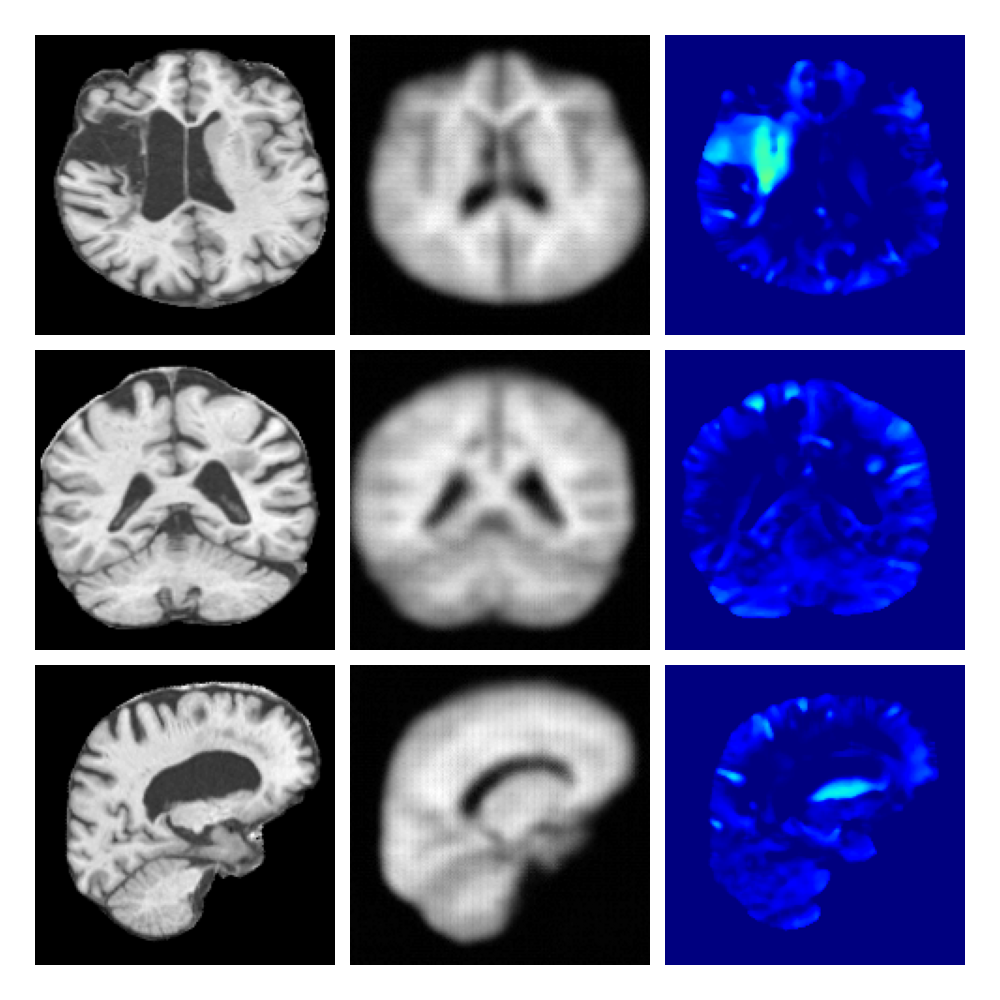}
        \caption{VAE}
    \end{subfigure}
    \begin{subfigure}{0.24\linewidth}
        \centering
        \includegraphics[width=\linewidth, trim=245 5 5 0, clip]{figures/sub-074_LDM.png}
        \caption{LDM}
    \end{subfigure}
    \begin{subfigure}{0.24\linewidth}
        \centering
         \includegraphics[width=\linewidth, trim=245 5 5 0, clip]{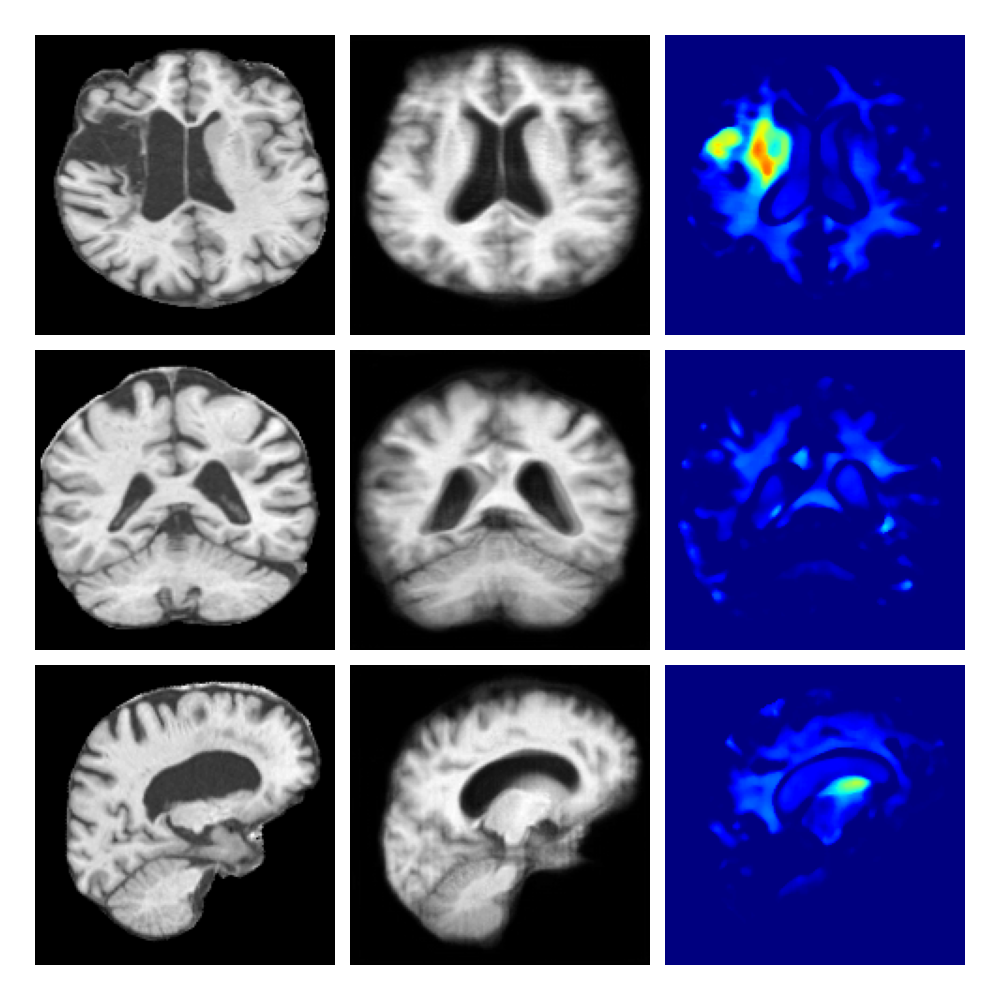}
        \caption{LDM ($T_{\text{avg}}$)}
    \end{subfigure}
    \begin{subfigure}{0.24\linewidth}
        \centering
        \includegraphics[width=\linewidth, trim=245 5 5 0, clip]{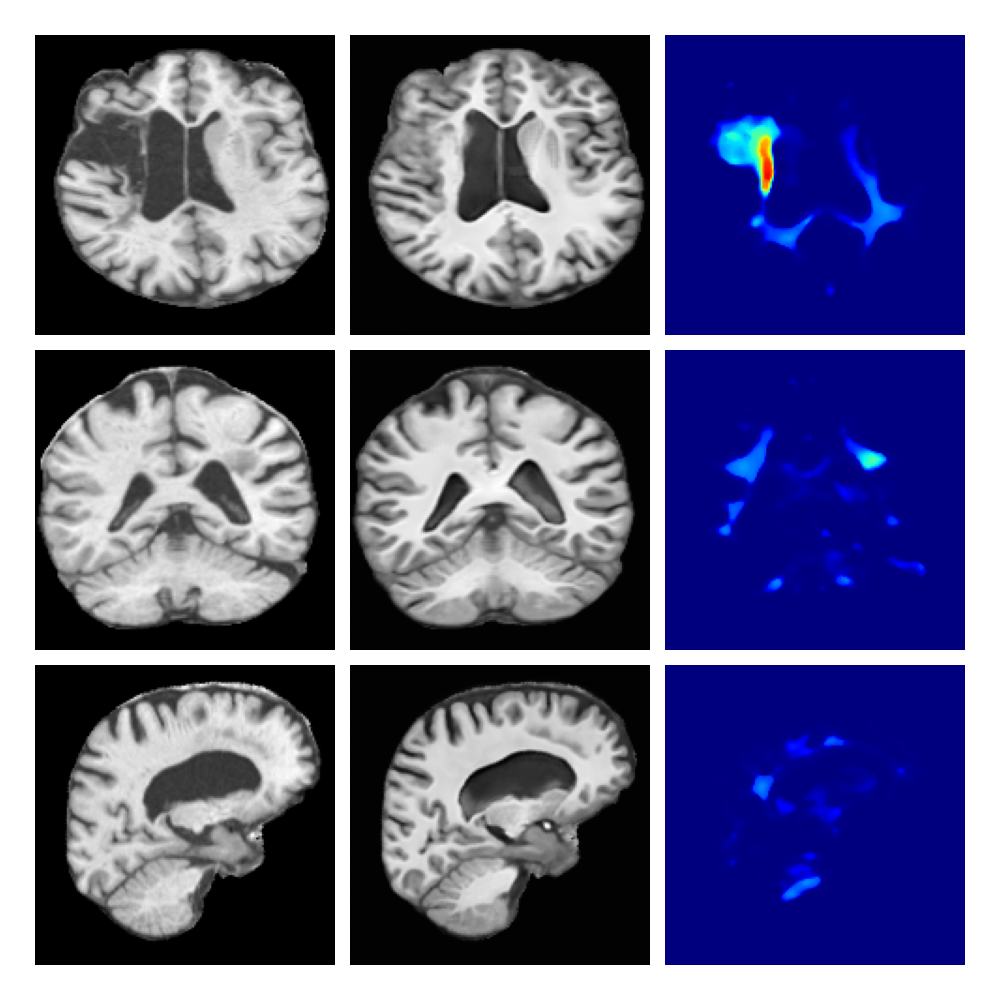}
        \caption{UNA}
    \end{subfigure}
    \begin{subfigure}{0.24\linewidth}
        \centering
        \includegraphics[width=\linewidth, trim=245 5 5 0, clip]{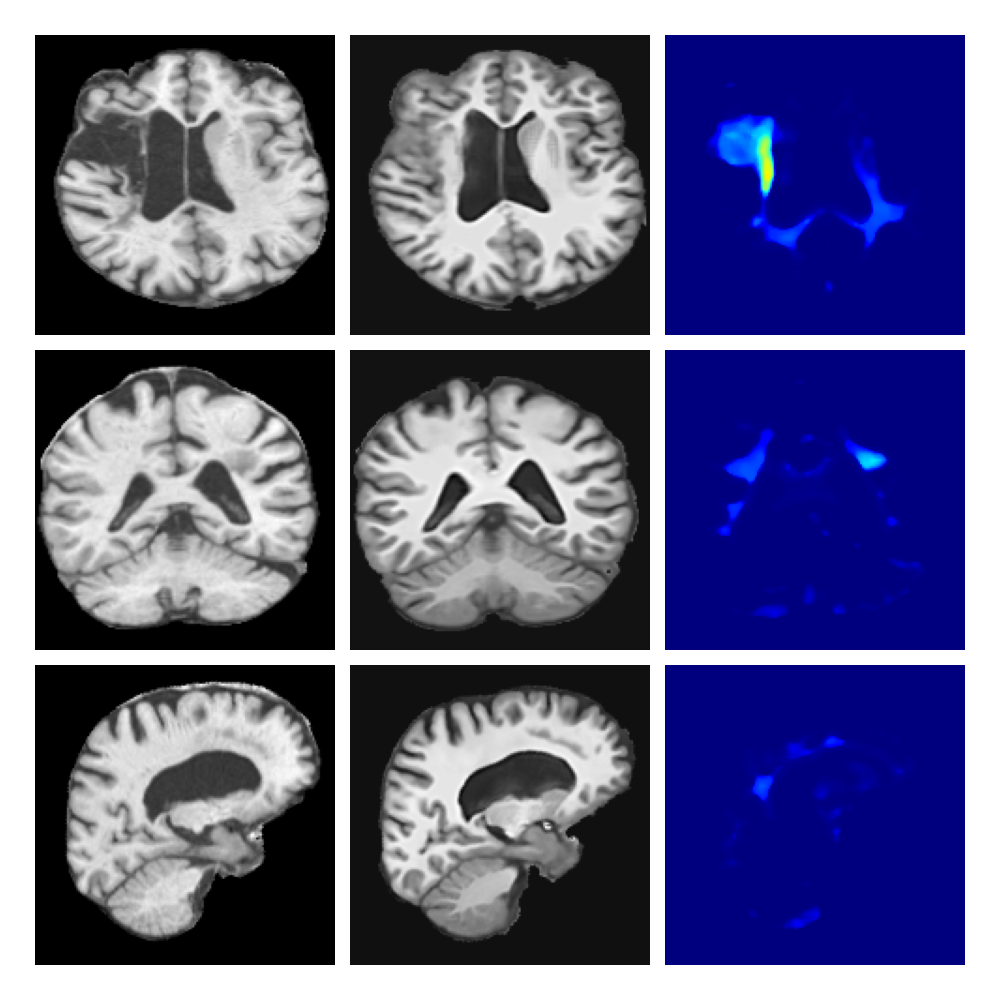}
        \caption{SynthSR}
    \end{subfigure}
    \begin{subfigure}{0.24\linewidth}
        \centering
          \includegraphics[width=\linewidth, trim=245 5 5 0, clip]{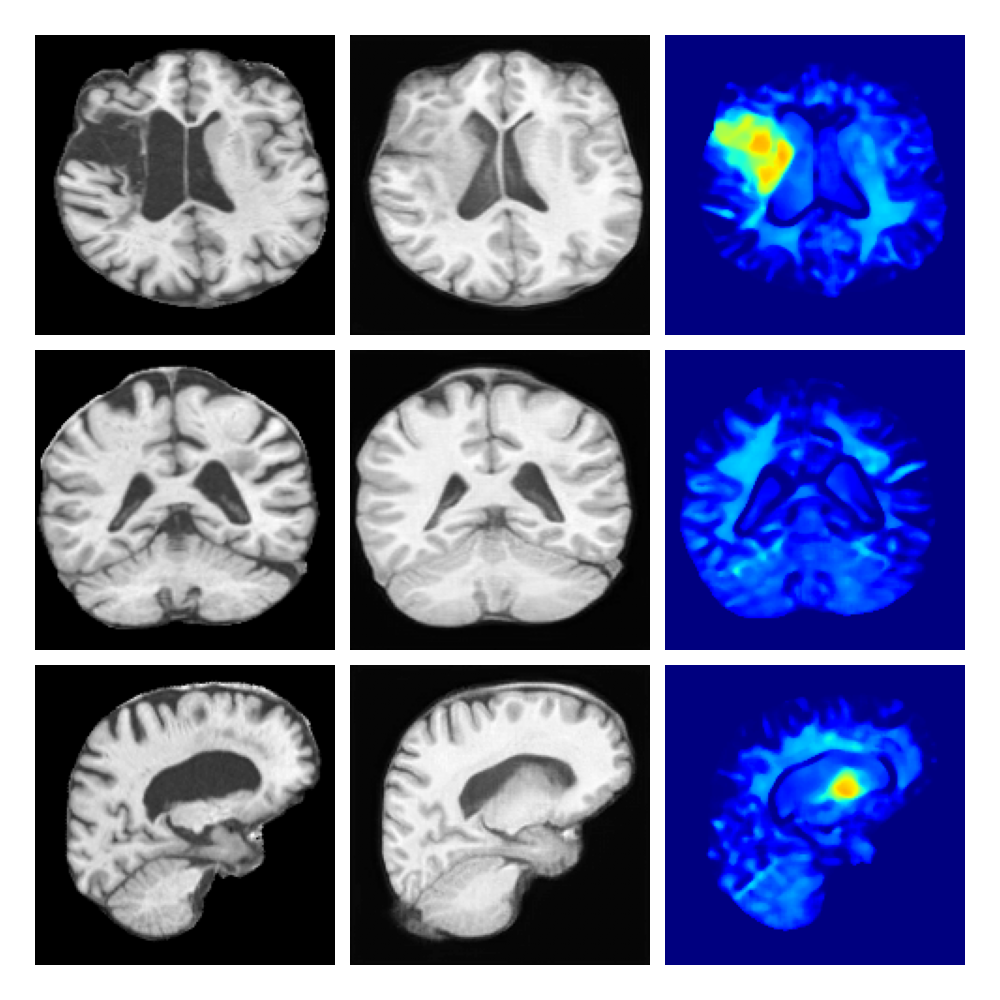}
        \caption{cLDM}
    \end{subfigure}
    \begin{subfigure}{0.24\linewidth}
        \centering
        \includegraphics[width=\linewidth, trim=245 5 5 0, clip]{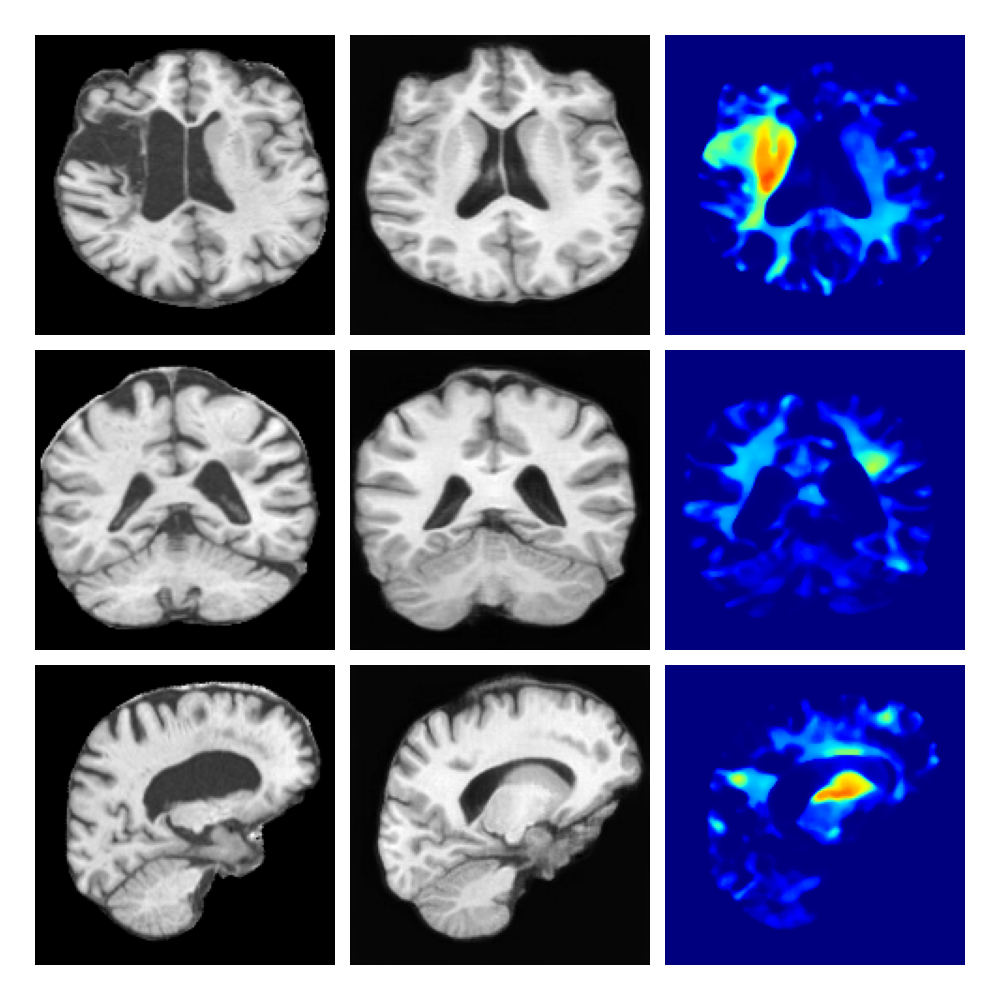}
        \caption{Ours}
    \end{subfigure}
    \caption{Example reconstructions and anomaly maps for an ATLAS test sample.} 
    \label{fig:pathology_results}
\end{figure*}

\section{Results}

Table \ref{tab:pathology_results} shows the quantitative pathology detection results. Our proposed model outperforms competing unsupervised approaches on every dataset and metric, and achieves a level of accuracy on par with SynthSR -- a fully supervised approach. Our method also outperforms SynthSR in FPR, highlighting our method's effectiveness in reducing false positive predictions and preserving individual characteristics. For real pathology from the ATLAS dataset, our method shows improvements of 14.72 \%, 12.26 \%, and 9.00 \% in Dice, $AP_{\text{pix}}$  and FPR, respectively, compared to the second-best method. UNA outperforms by 1.34 \% in $AUC_{\text{pix}}$. Our method achieves the best rank in 3 out of 5 synthetic datasets with SynthSR performing better on the HCP and ADHD200 cohorts. UNA achieves competitive performance, outperforming unsupervised methods on all datasets and SynthSR on the ATLAS cohort. Although cLDM shows improved performance over LDM, it is still substantially outperformed by our method. This suggests that our conditioning approach more effectively integrates pathological information, thus better guiding the reconstruction of pseudo-healthy images. 

Qualitatively, the VAE generates smooth reconstructions as expected (Figure \ref{fig:pathology_results}), leading to more false positives in healthy tissue. This corroborates the poor performance in the anomaly detection metrics in Table \ref{tab:pathology_results}. While SynthSR effectively preserves information from healthy tissue, Figure \ref{fig:pathology_results} shows that it sometimes fails to fully inpaint pathological regions -- particularly when these are large. Additionally, the SynthSR inpainted tissue does not appear anatomically realistic, which could limit the applicability of the pseudo-healthy images for downstream tasks. In contrast, our method produces a higher contrast anomaly map over abnormal regions, indicating superior localization of anomalies. While UNA achieves slightly better anomaly localization compared to SynthSR, it still falls short compared to our model. Although all diffusion-based methods successfully inpaint anomalous regions with realistic healthy tissue to some extent, they again do not achieve the same level of anomaly localization as our method.

\begin{table*}[ht!]
    \centering
    \smaller
    \caption{Anomaly detection metrics calculated using anomaly maps and manual annotations. \textbf{Bold} and \underline{underlined} indicate the best and second-best results. (Unsup.=Unsupervised, Sup.=Supervised, Weakly Sup.=Weakly Supervised.)}
    \label{tab:pathology_results}
    \begin{tabularx}{1.\textwidth}{
        p{1.2cm}p{1.3cm}
        X
        p{2cm}
        >{\centering\arraybackslash}p{1cm}
        >{\centering\arraybackslash}p{1.2cm}
        >{\centering\arraybackslash}p{1.5cm}
        >{\centering\arraybackslash}p{1.1cm}
        >{\centering\arraybackslash}p{1.1cm}
    }
        \toprule
        Source & Dataset & Type & Method & Dice $(\uparrow)$  & $AP_{\text{pix}}$ $(\uparrow)$ & $AUC_{\text{pix}}$ $(\uparrow)$ & FPR $(\downarrow)$ & Rank $(\downarrow)$\\
        \midrule
        \multirow{38}{*}{\parbox{2cm}{Synthetic\\pathology\\images}} 
        & \multirow{8}{*}{HCP} & \multirow{3}{*}{Unsup.} & VAE~\cite{Baur2020} & 0.2600 & 0.0144 & 0.7762 & 0.3304 & 6.00 \\
        & & & LDM & 0.1841 & 0.1391 & 0.8395 & 0.0562 & 5.75  \\
        & & & LDM ($T_{\text{avg}}$)~\cite{Graham2023b} & 0.1635 & 0.1349 & 0.8977 & 0.0530 & 5.75 \\
        & & \multirow{2}{*}{Sup.} & UNA~\cite{Liu2025} & 0.2525 & 0.2009 & \underline{0.9351} & 0.0278 & 3.00 \\
        & & & SynthSR~\cite{Iglesias2023} & \textbf{0.4151} & \textbf{0.3671} & \textbf{0.9450} & \underline{0.0193} & \textbf{1.00} \\
        & & \multirow{2}{*}{\parbox{1cm}{Weakly\\Sup.}} & cLDM & 0.2211 & 0.1920 & 0.9145 & 0.0340 & 4.25 \\
        & & & Ours & \underline{0.3511} & \underline{0.3159} & 0.9315 & \textbf{0.0130} & \underline{2.25} \\

        \cmidrule{2-9}
        & \multirow{8}{*}{ADNI} & \multirow{3}{*}{Unsup.} & VAE~\cite{Baur2020} & 0.0904 & 0.0077 & 0.7539 & 0.2716 & 7.00 \\  
        & & & LDM & 0.1395 & 0.0972 & 0.8154 & \underline{0.0252} & 5.00 \\
        & & & LDM ($T_{\text{avg}}$)~\cite{Graham2023b} & 0.1397 & 0.1120 & 0.9018 & 0.0563 & 5.00 \\
        & & \multirow{2}{*}{Sup.} & UNA~\cite{Liu2025} & 0.1502 & 0.1035 & \underline{0.9308} & 0.0404 & 4.00 \\
        & & & SynthSR~\cite{Iglesias2023} & \textbf{0.2553} & \textbf{0.2205} & 0.9297 & 0.0257 & \underline{2.00} \\
        
        & & \multirow{2}{*}{\parbox{1cm}{Weakly\\Sup.}} & cLDM & 0.1829 & 0.1568 & 0.9191 & 0.0341 & 3.25 \\
        & & & Ours & \underline{0.2493} & \underline{0.2050} & \textbf{0.9392} & \textbf{0.0117} & \textbf{1.50} \\
        \cmidrule{2-9}
        & \multirow{8}{*}{ADHD200}  & \multirow{3}{*}{Unsup.} & VAE~\cite{Baur2020} & 0.1472 & 0.0094 & 0.7472 & 0.3209 & 6.75 \\  
        & & & LDM & 0.1502 & 0.1032 & 0.8134 & 0.0395 & 5.25 \\
        & & & LDM ($T_{\text{avg}}$)~\cite{Graham2023b} & 0.1445 & 0.1183 & 0.8751 & 0.0831 &  5.75 \\
        
        & & \multirow{2}{*}{Sup.} & UNA~\cite{Liu2025} & 0.2324 & 0.1875 & \textbf{0.9363} & 0.0207 & 2.50 \\
        & & & SynthSR~\cite{Iglesias2023} & \textbf{0.3326} & \textbf{0.3056} & \underline{0.9313} & \underline{0.0193} & \textbf{1.50} \\
        
        & & \multirow{2}{*}{\parbox{1cm}{Weakly\\Sup.}} & cLDM & 0.1953 & 0.1697 & 0.8933 & 0.0543 & 4.25 \\
        & & & Ours & \underline{0.3105} & \underline{0.2688} & 0.9235 & \textbf{0.0175} & \underline{2.00} \\
        \cmidrule{2-9}
        & \multirow{8}{*}{OASIS3} & \multirow{3}{*}{Unsup.} & VAE~\cite{Baur2020} & 0.0988 & 0.0084 & 0.7551 & 0.2826 & 7.00 \\  
        & & & LDM & 0.1315 & 0.0902 & 0.8124 & 0.0407 & 5.75 \\
        & & & LDM ($T_{\text{avg}}$)~\cite{Graham2023b} & 0.1365 & 0.1099 & 0.8872 & 0.0756 & 5.25 \\
        & & \multirow{2}{*}{Sup.} & UNA~\cite{Liu2025} & 0.1655 & 0.1116 & \underline{0.9344} & 0.0303 & 3.25 \\
        & & &  SynthSR~\cite{Iglesias2023} & \textbf{0.2712} & \textbf{0.2363} & 0.9342 & \underline{0.0224} & \underline{1.75} \\
        & & \multirow{2}{*}{\parbox{1cm}{Weakly\\Sup.}} & cLDM & 0.1925 & 0.1646 & 0.9096 & 0.0400 & 3.50 \\
        & & & Ours & \underline{0.2666} & \underline{0.2288} & \textbf{0.9418} & \textbf{0.0130} & \textbf{1.50} \\
        \cmidrule{2-9}
        & \multirow{8}{*}{AIBL} & \multirow{3}{*}{Unsup.} & VAE~\cite{Baur2020} & 0.1092 & 0.0083 & 0.7639 & 0.2676 & 7.00 \\  
        & & & LDM & 0.1617 & 0.1148 & 0.8317 & 0.0261 & 4.50 \\
        & & & LDM ($T_{\text{avg}}$)~\cite{Graham2023b} & 0.1397 & 0.1120 & 0.9018 & 0.0563 & 5.75 \\
        & & \multirow{2}{*}{Sup.} & UNA~\cite{Liu2025} & 0.1629 & 0.1122 & 0.9319 & 0.0315 & 4.00 \\
         & & & SynthSR~\cite{Iglesias2023} & \underline{0.2725} & \underline{0.2339} & \underline{0.9349} & \underline{0.0193} & \underline{2.00} \\
        & & \multirow{2}{*}{\parbox{1cm}{Weakly\\Sup.}} & cLDM & 0.2211 & 0.1920 & 0.9145 & 0.0340 & 3.75 \\
        & & & Ours & \textbf{0.2858} & \textbf{0.2431} & \textbf{0.9418} & \textbf{0.0130} & \textbf{1.00} \\
        \midrule
        \multirow{8}{*}{\parbox{1.4cm}{Real\\stroke\\images}} 
        & \multirow{8}{*}{ATLAS} & \multirow{3}{*}{Unsup.} & VAE~\cite{Baur2020} & 0.1646 & 0.1341 & 0.8921 & 0.0506 & 7.00 \\ 
        & & & LDM & 0.2418 & 0.1889 & 0.9303 & 0.0124 & 4.25 \\
        & & & LDM ($T_{\text{avg}}$)~\cite{Graham2023b} & 0.2018 & 0.1567 & 0.9037 & 0.0495 & 6.00 \\
        & & \multirow{2}{*}{Sup.} & UNA~\cite{Liu2025} & 0.2614 & 0.2210 & \textbf{0.9539} & \underline{0.0100} & \underline{2.25} \\
        & & & SynthSR~\cite{Iglesias2023} & \underline{0.2711} & \underline{0.2354} & \underline{0.9367} & 0.0136 & 2.75 \\
        & & \multirow{2}{*}{\parbox{1cm}{Weakly\\Sup.}} & cLDM & 0.2588 & 0.2091 & 0.9236 &  0.0146 & 4.50 \\
        & & & Ours & \textbf{0.3110} & \textbf{0.2643} & \underline{0.9413} & \textbf{0.0091} & \textbf{1.25} \\
        \bottomrule
    \end{tabularx}
\end{table*}

\section{Conclusion and further work}

In this work, we introduced a novel pseudo-disease conditioned diffusion model framework for anomaly detection in 3D pathological T1 MRI. Our model achieves state-of-the-art performance across both synthetic and real pathology datasets. The high Dice scores and low FPR in Table \ref{tab:pathology_results} (on par with those of a supervised approach), coupled with the high contrast anomaly map in Figure \ref{fig:pathology_results}, demonstrate that our model effectively inpaints pathological tissue while preserving individual-level characteristics. As such, our approach to incorporating pathological images into the diffusion model training process improves the inpainting of diseased regions while preventing the loss of critical information from surrounding healthy tissue. Further work will involve applying the generated pseudo-healthy reconstructions to other downstream tasks, such as brain parcellation or tissue segmentation. Our method holds great promise in analyzing large uncurated clinical datasets with heterogeneous types of pathology.

%
%
\newpage
\bibliographystyle{splncs04}
\bibliography{main}
\end{document}